\documentclass[aps,prl,twocolumn,superscriptaddress,showpacs,preprintnumbers,amsmath,amssymb]{revtex4-1}

\usepackage{amsmath}

\usepackage{subfigure}
\def\am{$^{241}$Am~}

\def\pb{$^{210}$Pb~}

\def\cenns{CE$\nu$NS~}

\usepackage{romanbar}
\usepackage{xcolor}
\usepackage{graphicx}
\usepackage{dcolumn}
\usepackage{bm}

\usepackage{array}
\usepackage{makecell}

\begin{document}

\title{Upgrade of NaI(Tl) crystal encapsulation for the NEON experiment }

\author{J.~J.~Choi}\affiliation{Department of Physics and Astronomy, Seoul National University, Seoul 08826, Republic of Korea} 
\author{E.~J.~Jeon}\affiliation{Center for Underground Physics, Institute for Basic Science (IBS), Daejeon 34126, Republic of Korea}\affiliation{IBS School, University of Science and Technology (UST), Deajeon 34113, Republic of Korea}
\author{J.~Y.~Kim}\affiliation{Korea Atomic Energy Research Institute, Deajeon 34057, Republic of Korea} 
\author{K.~W.~Kim}\affiliation{Center for Underground Physics, Institute for Basic Science (IBS), Daejeon 34126, Republic of Korea}
\author{S.~H.~Kim}\affiliation{Center for Underground Physics, Institute for Basic Science (IBS), Daejeon 34126, Republic of Korea}
\author{S.~K.~Kim}\affiliation{Department of Physics and Astronomy, Seoul National University, Seoul 08826, Republic of Korea} 
\author{Y.~D.~Kim}\affiliation{Center for Underground Physics, Institute for Basic Science (IBS), Daejeon 34126, Republic of Korea}\affiliation{IBS School, University of Science and Technology (UST), Deajeon 34113, Republic of Korea}
\author{Y.~J.~Ko}\affiliation{Center for Underground Physics, Institute for Basic Science (IBS), Daejeon 34126, Republic of Korea}
\author{B.~C.~Koh}\affiliation{Department of Physics, Chung-Ang University, Seoul 06973, Republic of Korea}

\author{C.~Ha}\affiliation{Department of Physics, Chung-Ang University, Seoul 06973, Republic of Korea}

\author{B.~J.~Park}\affiliation{Center for Underground Physics, Institute for Basic Science (IBS), Daejeon 34126, Republic of Korea}
\author{S.~H.~Lee}\affiliation{IBS School, University of Science and Technology (UST), Deajeon 34113, Republic of Korea}\affiliation{Center for Underground Physics, Institute for Basic Science (IBS), Daejeon 34126, Republic of Korea}
\author{I.~S.~Lee}\affiliation{Center for Underground Physics, Institute for Basic Science (IBS), Daejeon 34126, Republic of Korea}
\author{H.~Lee}\affiliation{IBS School, University of Science and Technology (UST), Deajeon 34113, Republic of Korea}\affiliation{Center for Underground Physics, Institute for Basic Science (IBS), Daejeon 34126, Republic of Korea}
\author{H.~S.~Lee}\affiliation{Center for Underground Physics, Institute for Basic Science (IBS), Daejeon 34126, Republic of Korea}\affiliation{IBS School, University of Science and Technology (UST), Deajeon 34113, Republic of Korea}
\author{J.~Lee}\affiliation{Center for Underground Physics, Institute for Basic Science (IBS), Daejeon 34126, Republic of Korea}
\author{Y.~M.~Oh} \affiliation{Center for Underground Physics, Institute for Basic Science (IBS), Daejeon 34126, Republic of Korea}
\collaboration{(NEON Collaboration)}

\affiliation{Department of Physics and Astronomy, Seoul National University, Seoul 08826, Republic of Korea} 
\affiliation{Center for Underground Physics, Institute for Basic Science (IBS), Daejeon 34126, Republic of Korea}
\affiliation{IBS School, University of Science and Technology (UST), Deajeon 34113, Republic of Korea}
\affiliation{Korea Atomic Energy Research Institute, Deajeon 34057, Republic of Korea} 
\affiliation{Department of Physics, Chung-Ang University, Seoul 06973, Republic of Korea}

\begin{abstract}
  
The Neutrino Elastic-scattering Observation with NaI(Tl) experiment (NEON) aims to detect coherent elastic neutrino-nucleus scattering~(\cenns) in a NaI(Tl) crystal using reactor anti-electron neutrinos at the Hanbit nuclear power plant complex.
A total of 13.3 kg of NaI(Tl) crystals were initially installed in December 2020 at the tendon gallery, 23.7$\pm$0.3\,m away from the reactor core, which operates at a thermal power of 2.8\,GW.
Initial engineering operation was performed from May 2021 to March 2022 and observed unexpected photomultiplier-induced noise and a decreased light yield that were induced by leakage of liquid scintillator into the detector due to weakness of detector encapsulation.
We upgraded the detector encapsulation design to prevent leakage of the liquid scintillator. Meanwhile two small size detectors were replaced with large size detectors returning a total mass of 16.7\,kg. 
With this new design implementation, the detector system has been operating stably since April 2022 for over a year without detector gain drop. 
In this paper, we present an improved crystal encapsulation design and stability of the NEON experiment.  
\end{abstract}

\maketitle

\flushbottom

\section{Introduction}
Although coherent elastic neutrino-nucleus scattering~(\cenns) was predicted by the Standard Model in 1974~\cite{PhysRevD.9.1389,osti_4289450}, its first observation took more than 40 years. This milestone was achieved by the COHERENT collaboration using a CsI(Na) detector in a stopped-pion source from the SNS accelerator in 2017~\cite{Akimov:2017ade}. Following this, the same research team verified the process using a liquid argon detector~\cite{COHERENT:2020iec} and germanium detectors~\cite{COHERENT:Ge}. The resulting signal includes, in part, the incoherent scattering at the neutrino energy of $\sim$30\,MeV~\cite{Kerman:2016jqp}. Therefore, investigation of \cenns in the fully coherent regime using the few MeV antineutrinos from a nuclear reactor can provide precise information for understanding neutrinos. This \cenns represents a crucial avenue for testing Standard Model~\cite{Drukier19842295,Krauss:1991ba,Patton:2012jr}, exploring non-standard neutrino interactions~\cite{Formaggio:2011jt,deNiverville:2015mwa,Dutta:2015nlo,Dent:2016wcr,Kosmas:2017zbh,Liao:2017uzy,Farzan:2018gtr,Dev:2019anc,Aguilar-Arevalo:2019jlr}, modeling supernova energy transport in astrophysics~\cite{Janka:2017vlw}, and monitoring nuclear reactors~\cite{Cogswell2016,RevModPhys.92.011003}. Several experiments are currently running or under construction to detect \cenns from reactors such as CONUS~\cite{CONUS:2020skt}, NUCLEUS~\cite{Angloher:2019flc}, CONNIE~\cite{Aguilar-Arevalo:2016khx}, and $\nu$GeN~\cite{Belov:2015ufh}.



 Neutrino Elastic-scattering Observation with NaI (NEON) aims to detect \cenns using reactor antineutrinos from reactor unit 6 at the Hanbit nuclear power plant complex~\cite{Choi_2023}.

NaI(Tl) crystals for the NEON detector employ a novel technique of crystal encapsulation that significantly increased the light collection efficiency, achieving approximately 22 photoelectrons (PEs) per keV electron-equivalent energy (keVee)~\cite{CHOI2020164556, Choi_2023}, which is approximately 50\,$\%$ larger than the light yield of COSINE-100 crystals~\cite{Adhikari:2017esn}.
The six NaI(Tl) crystal detector are immersed in 800 L of liquid scintillator (LS) to actively reduce radioactive backgrounds~\cite{Adhikari:2020asl}. The LS is surrounded by 10\,cm lead, 2.5\,cm borated-polyethylene, and 20\,cm polyethylene to reduce the background contribution form external radiation~\cite{Choi_2023}.
 
 To monitor the initial performance and detector stability, engineering operations were performed from May 2021 to Mar 2022 using 13.3\,kg of NaI(Tl) crystals. 
 During the engineering run, we found that the crystal light yield gradually decreased and unexpected photomultiplier (PMT)-induced noise occurred due to leakage of LS into the detector, resulting from weaknesses in the detector encapsulation. We discussed a conceptual design for an upgraded encapsulation to resolve this issue~\cite{Choi_2023}. 
 
 This paper presents new findings from the real assemblies and measurements conducted during the NEON experiment using the upgraded crystal encapsulation. 
 By implementing new encapsulation to the NEON crystals, we have significantly improved the operational stability and sensitivity of the NEON experiment. These advancements not only contribute to the NEON experiment but also provide valuable insights for future neutrino and dark matter search experiments utilizing NaI(Tl) crystals.
 


\section{NEON encapsulation design}

\begin{figure}[tbh!]
\begin{center}
\includegraphics[width=0.5\textwidth]{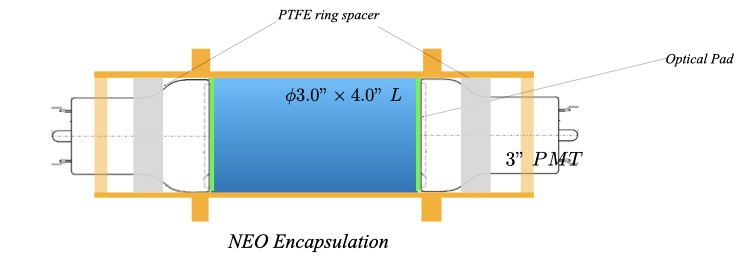} 
\caption{NaI(Tl) crystal encapsulation design used for the NEON engineering run}
\label{fig:phase1Encapsulation}
\end{center}
\end{figure}
The NEON experiment utilized six commercial-grade NaI(Tl) crystals manufactured by Alpha Spectra Inc. (AS). During the engineering run, a total of 13.3 kg of NaI crystals were employed: four with a 3-inch diameter and 4-inch length, and two with a 3-inch diameter and 8-inch length. 
To maximize the measured light yield while protecting the NaI(Tl) from moisture due to its hygroscopic properties, we applied a novel technique for crystal encapsulation to NEON detectors~\cite{CHOI2020164556}. 
This design minimizes the optical interface between the crystal and PMT by utilizing an optical pad (silicone rubber EJ-560 from Eljen technology) to efficiently collect scintillation photons. The crystals and PMTs are encapsulated within a copper case, exposing only the PMT pins to tightly seal the PMT necks, as illustrated in Fig.~\ref{fig:phase1Encapsulation}.

\begin{figure}[tbh!]
\begin{center}
      \subfigure[\label{fig:signalwaveform}]{\includegraphics[width=0.4\textwidth]{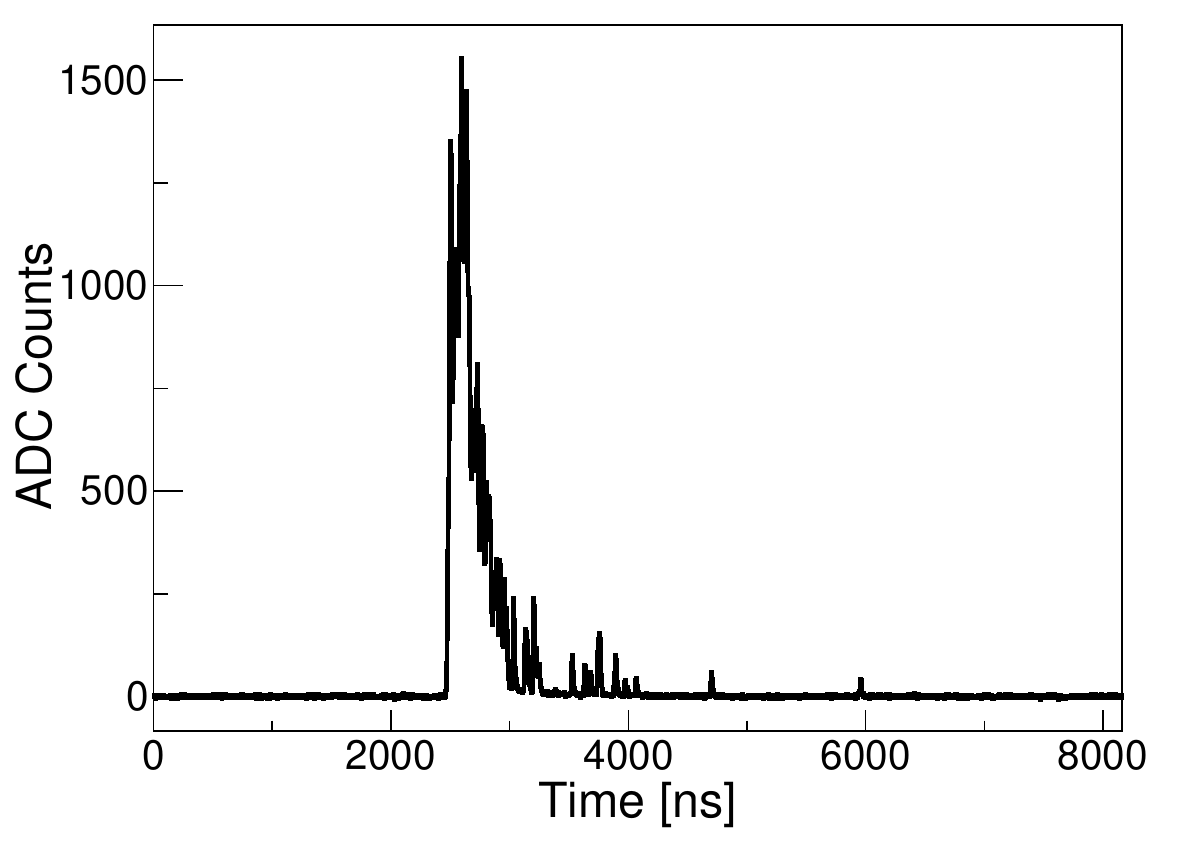}}
     \subfigure[\label{fig:noisewaveform}]{\includegraphics[width=0.4\textwidth]{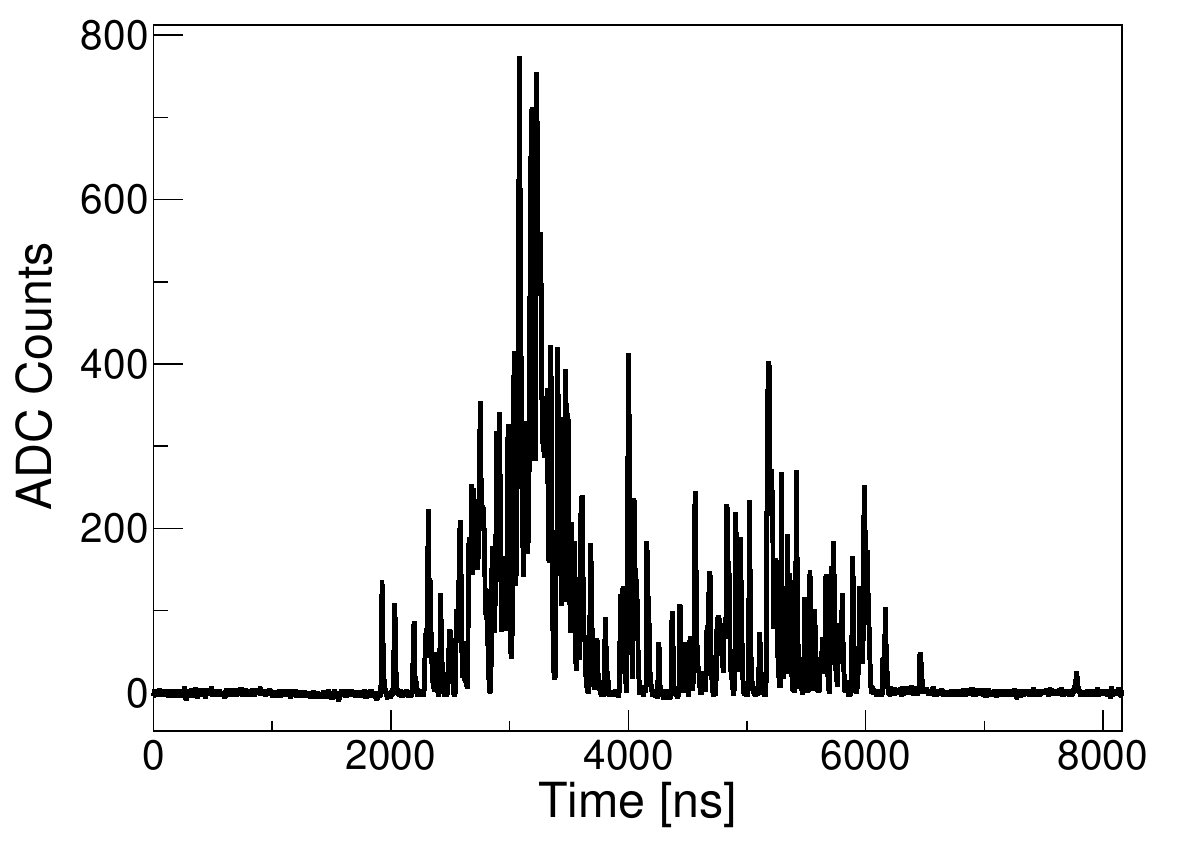}}
\caption{(a) A waveform of typical scintillation events of 20 keV in NaI(Tl) crystal. (b) An example of the PMT-induced noise events when PMT-base exposed to LS directly..}
\label{fig:GaussianNoise}
\end{center}
\end{figure}

\begin{figure}[tbh!]
\begin{center}
     \subfigure[\label{fig:meantime}]{\includegraphics[width=0.4\textwidth]{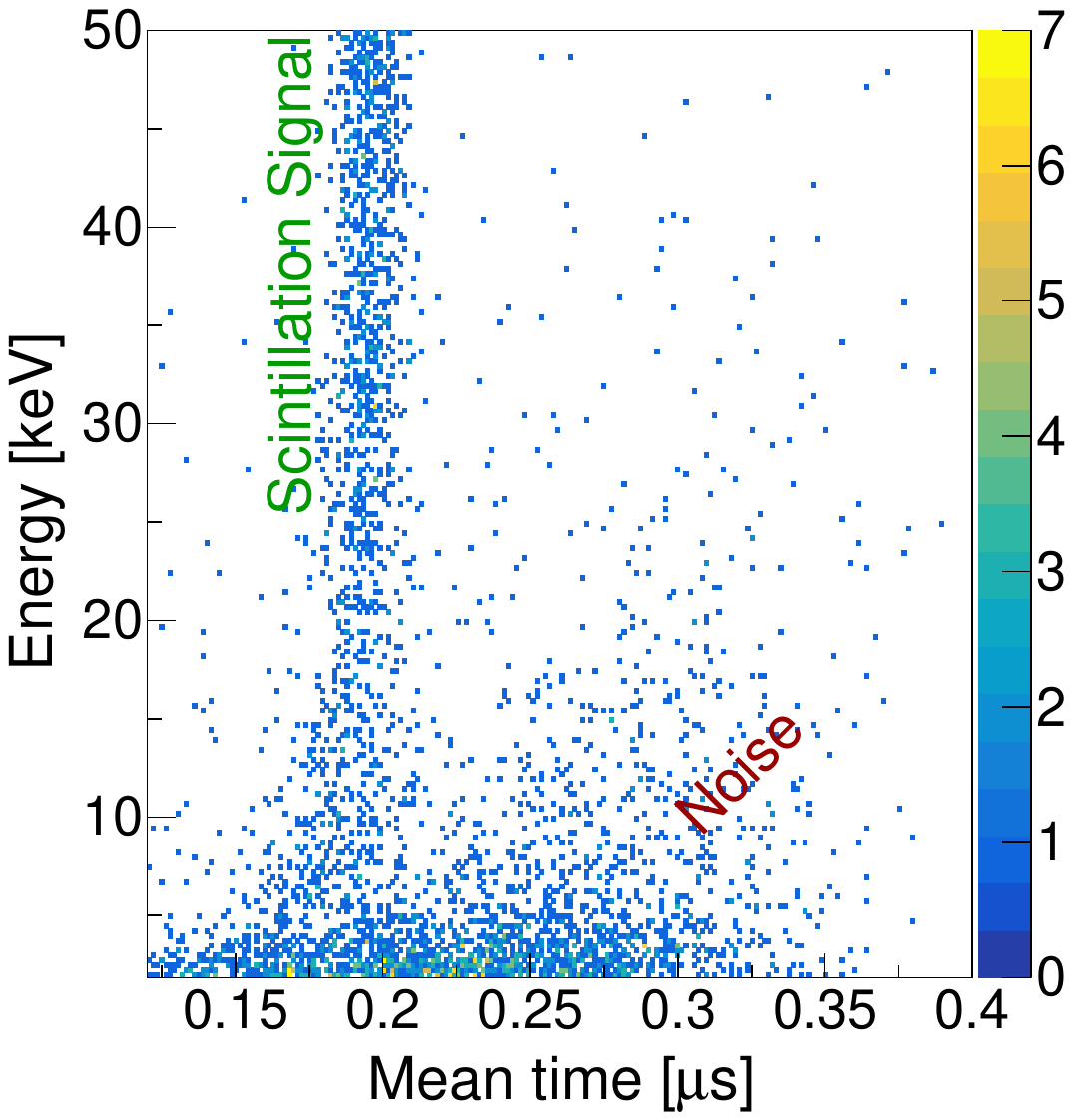}}  \centering
       \subfigure[\label{fig:meantime2}]{\includegraphics[width=0.4\textwidth]{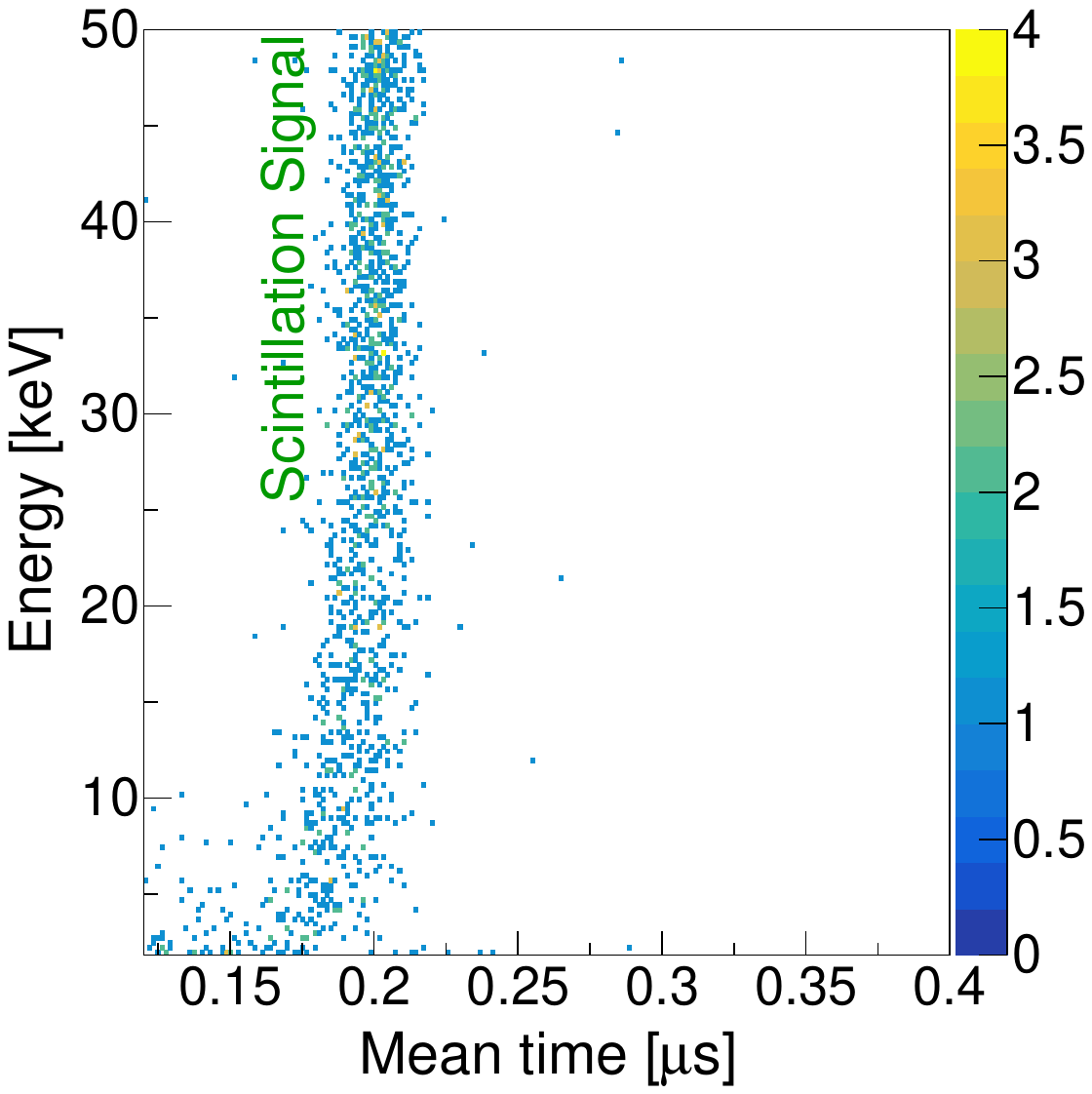}}  \centering
\caption{(a) The mean time distribution with the PMT-induced noise when PMT bases are exposed LS. (b) The mean time distribution after encapsulation upgrade.}
\label{fig:meantimeall}
\end{center}
\end{figure}

For active reduction in radiogenic backgrounds, the NEON detector is installed inside of linear alkylbenzene-based LS~\cite{Choi_2023}. However, this encapsulation design, where the PMT pins and PMT bases are exposed to LS, caused the unexpected PMT-induced noise distributed in the low-energy region. 
Figure~\ref{fig:noisewaveform} visually represents this unexpected noise as an example waveform. It is populated below 10\,keV but, broadly distributed up to 50\,keV. This differs from typical PMT-induced noise due to Cherenkov radiation in the PMT glass, which generate fast-decaying noise~\cite{Adhikari:2017esn, kwkim15}. Due to its broad range meantime distribution as shown in Fig.~\ref{fig:meantime}, these kinds of events are difficult to discriminate from  scintillation events, especially in the low-energy region below 5\,keV. We found that this PMT-induced noise occurred in all PMTs with the previous encapsulation design.

\begin{figure}[tbh!]
\begin{center}
    \includegraphics[width=0.5\textwidth]{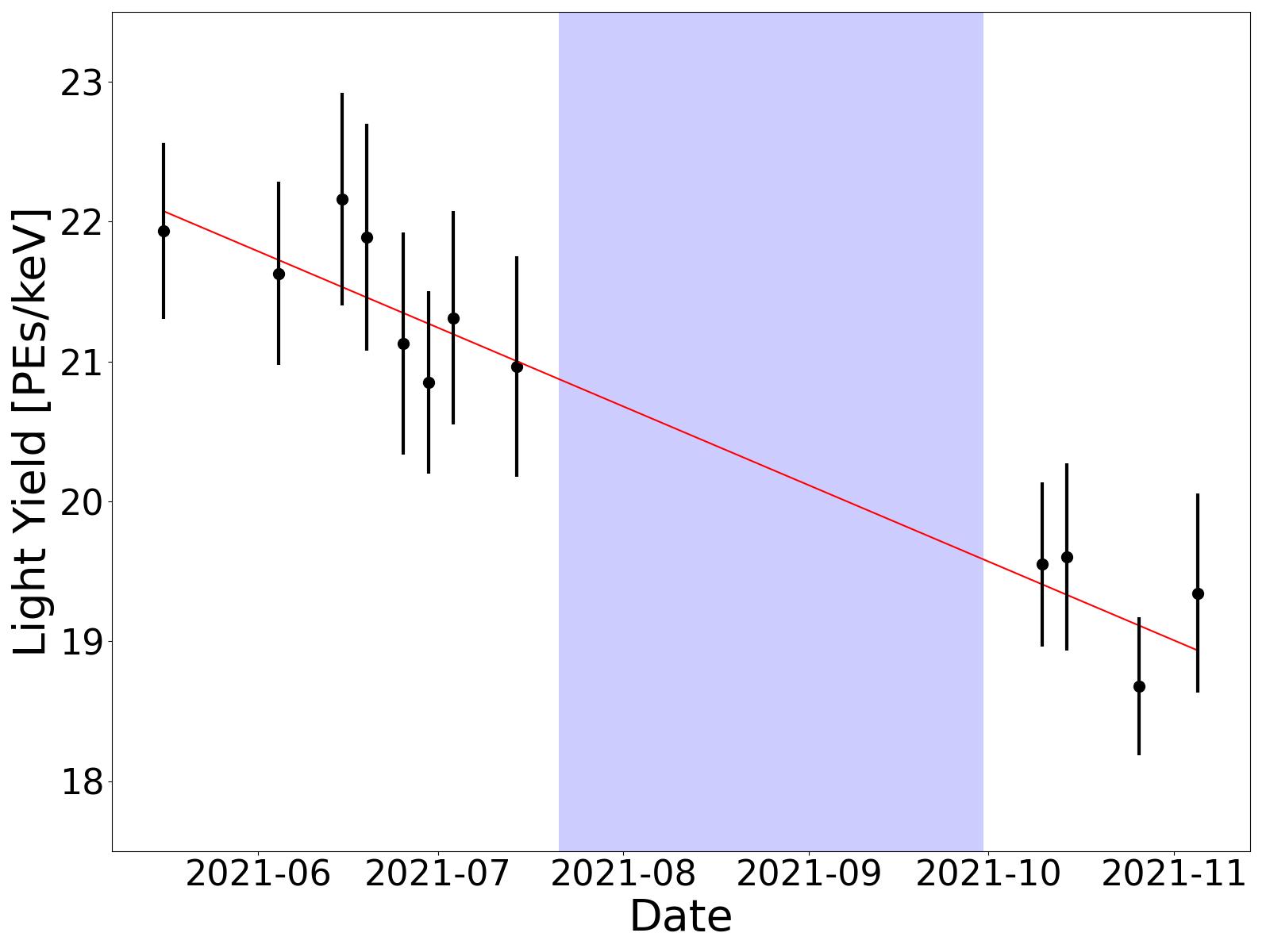}
    \caption{The DET-4 light yield monitoring while receiving engineering data. 
    The light yield calculation method is described in section~\ref{ch:LY} and didn't take into account low gain of photoelectrons.
    The colored time periods in the figure represent the time periods during which calibration data was received.
    During the time of receiving the engineering run, the light yield decreased over time, indicating that the encapsulation design used in the engineering run was not stable enough to maintain light yield. 
    }
\label{fig:Gaindrop}
\end{center}
\end{figure}

In addition to the unexpected PMT-induced noise events, we found that LS could penetrate into the encapsulation over a few months. Because of the glass structure of the PMT neck, the pressure to seal the PMT necks was limited. We also found that each PMT had slightly different geometric shape, making it hard to seal this part perfectly. Over time, LS continuously leaked  into the encapsulation and contaminated the reflector on the crystal surface. Smearing into the optical interface detached the optical coupling as well. Small white spots on the crystal surface due to humidity of LS was also developed. All this led to a reduced light yield over time, as shown in Fig.~\ref{fig:Gaindrop}. Similar decreases in light yield were observed in all other crystals, necessitating a new encapsulation design to protect against LS leak into the crystal and the PMT base~\cite{Choi_2023}.

\begin{figure}[!htb]
  \begin{center}
  \subfigure[\label{fig:inner}]{\includegraphics[width=0.45\textwidth]{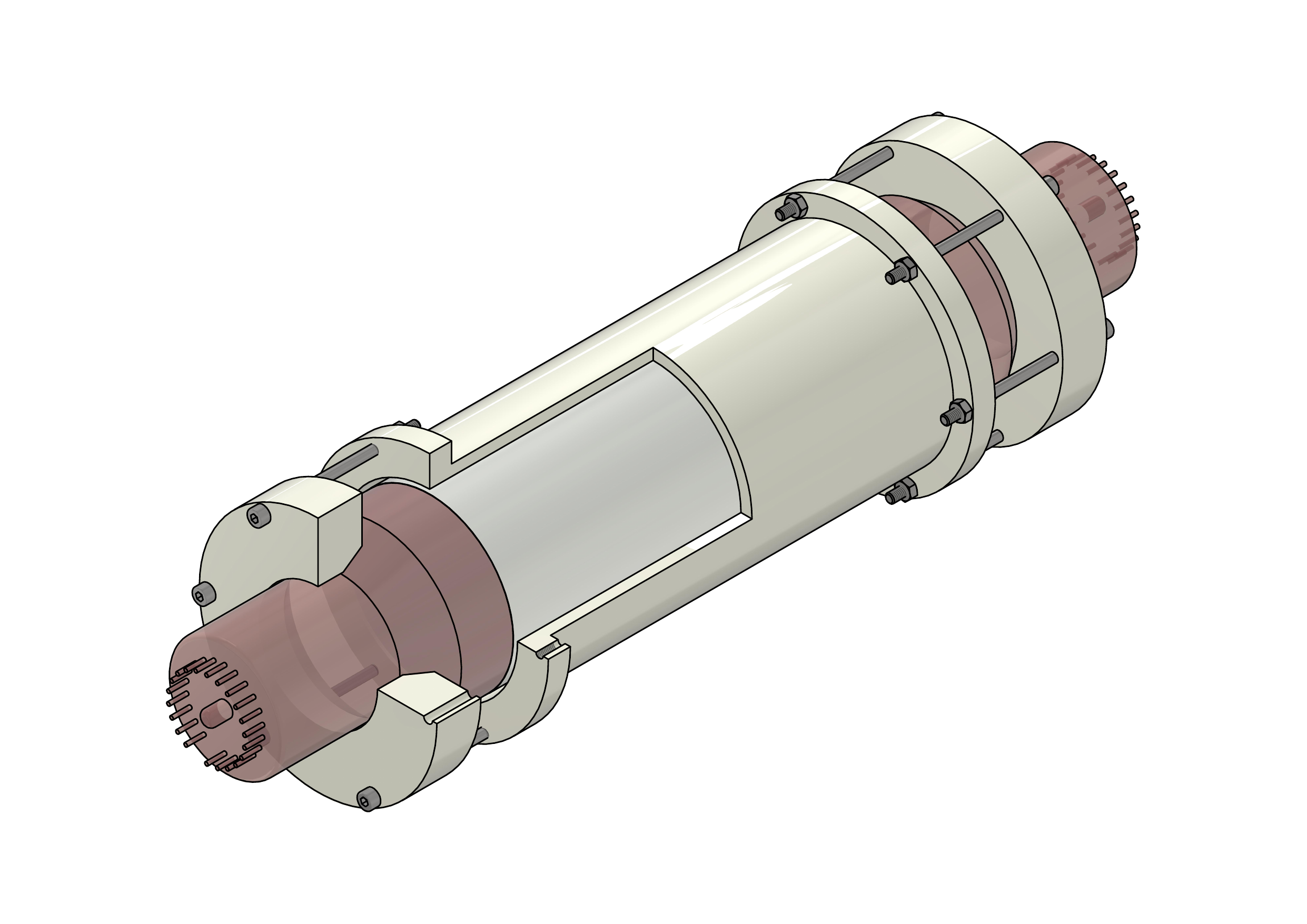}}
    \subfigure[\label{fig:case}]{\includegraphics[width=0.45\textwidth]{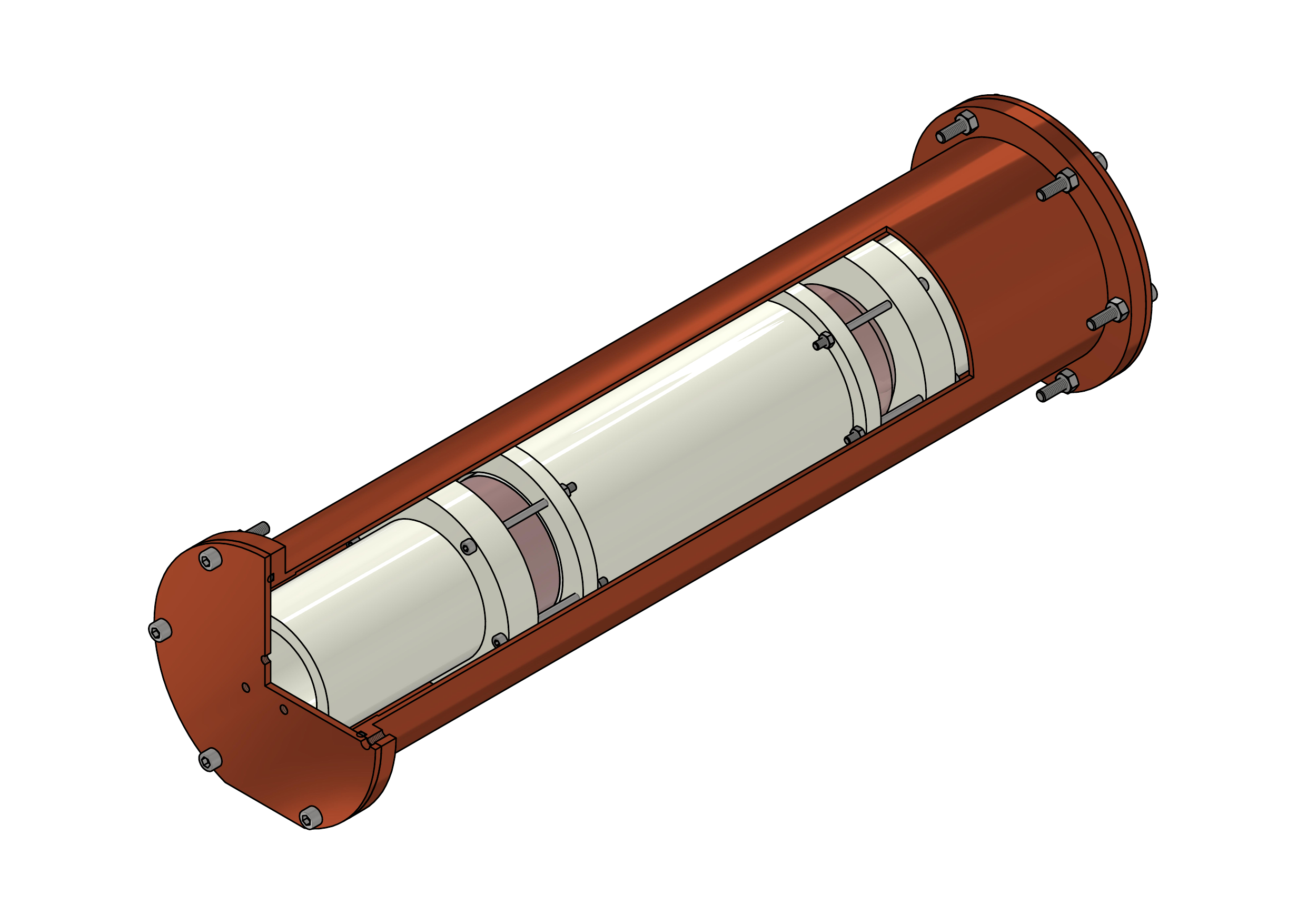}}
    
  \end{center}
  \caption{Updated design of the NaI(Tl) detector encapsulation. (a) Inner structure for a mounting between the crystal and PMTs. (b) Encasement of the crystal-PMTs structure in the copper case.}
  \label{fig:newencap}
\end{figure}

\begin{figure}[!htb]\centering
  
        \subfigure[\label{fig:teflonstructure}]{\includegraphics[width=0.4\textwidth]{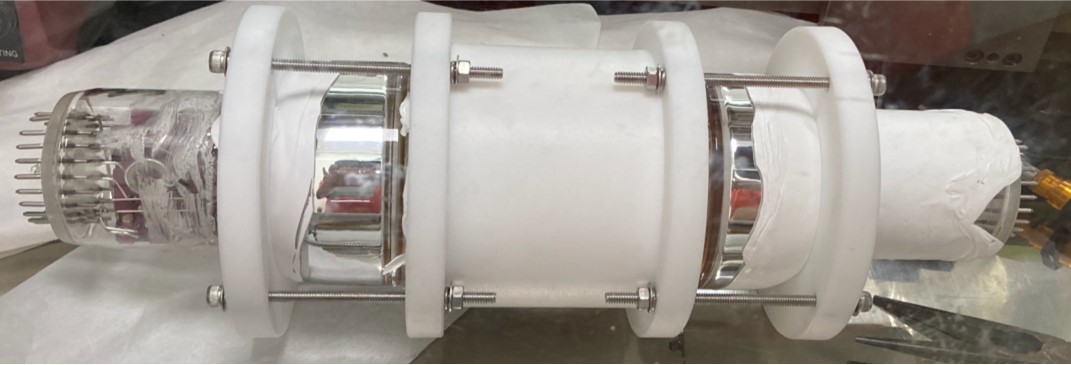} }
        \subfigure[\label{fig:coppercase}]{\includegraphics[height=0.18\textwidth]{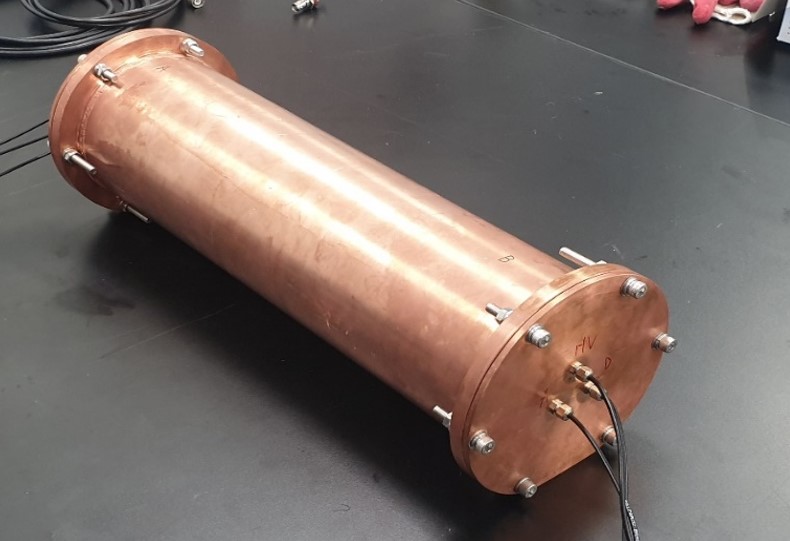} }
        \subfigure[\label{fig:cablegland}]{ \includegraphics[angle=90,height=0.18\textwidth]{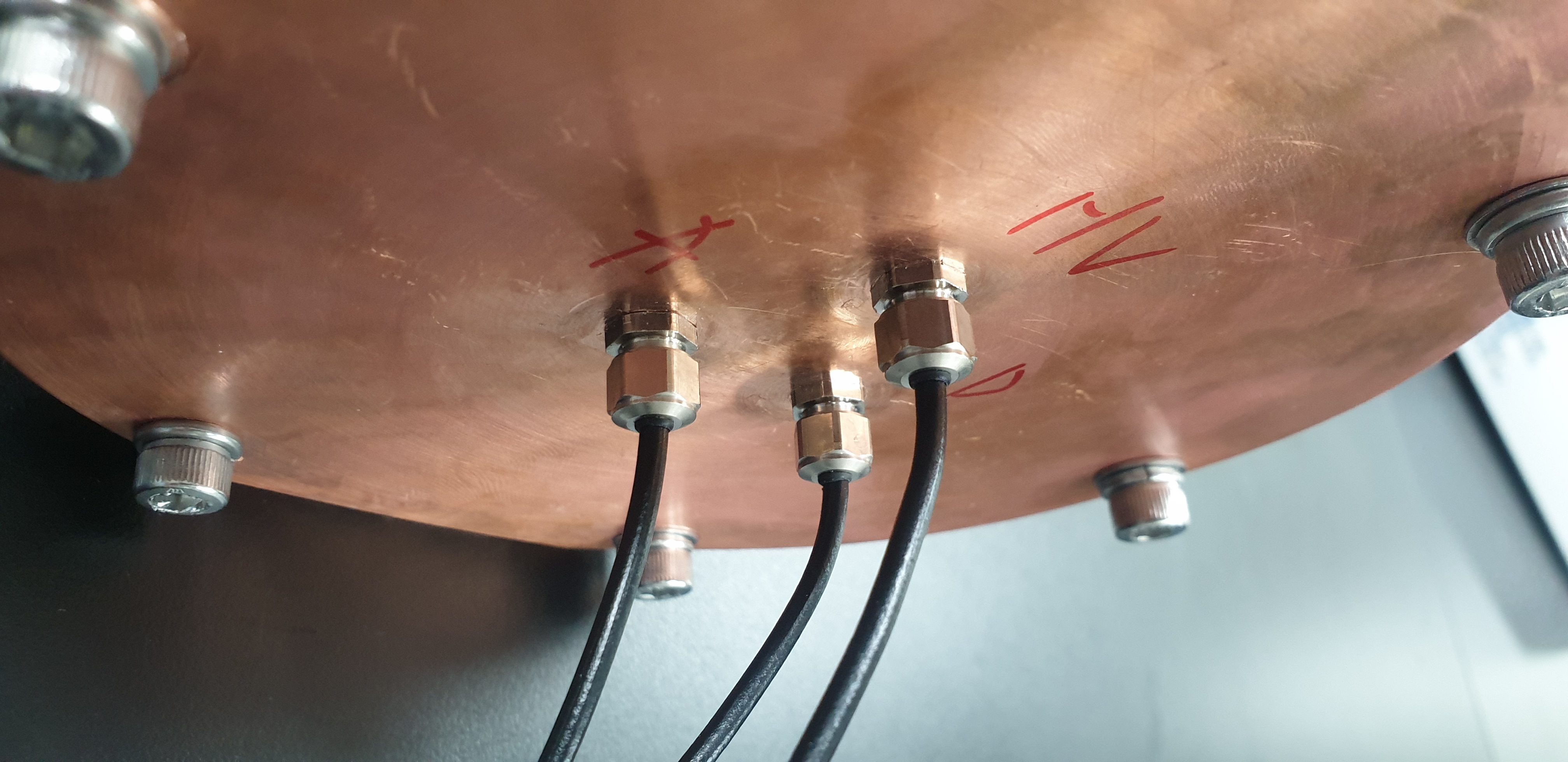}  }

  \caption{Picture of 4-inch length crystal assembly. (a) Inner structure of 4-inch long crystal and PMTs assembly. (b) Encapsulated detector with copper case and cable glands. (c) Zoomed picture of cable grands to secure detector.  } 
\label{fig:newdetector}
\end{figure}

The new encapsulation comprises two key components: an inner structure dedicated to maintaining a stable coupling between PMT and crystals, and a copper case designed to prevent the infiltration of outside air and LS.
A schematic design of the inner structure is shown in Fig.~\ref{fig:teflonstructure}.
The crystal detector with the previous encapsulation was disassembled in a glovebox, where the humidity was maintained at less than 10\,ppm (H$_{2}$O) using Ar gas and a molecular sieve trap. The crystal surface was cleaned with 99.9\% ethanol and to remove any remained LS and white spots. To remove humidity contamination in the ethanol, the surface treatment was finished with 99.9\% isopropyl alcohol. Optically interfacing two windows were dry-polished using aluminum oxide lapping films ranging from 400 to 800 grits and cleaned with isopropyl alcohol.     
After the polishing the crystal, several layers of soft Teflon sheets were wrapped around the barrel surface as a diffusive reflector, and the optical pad was coupled to the top and bottom surfaces. Meanwhile, we prepared all assembly parts by cleaning them in a diluted Citranox solution with sonication, baking them in a vacuum oven, and drying them in the glovebox for 2 days. 

The NaI(Tl) crystal is positioned within a polytetrafluoroethylene(PTFE) flanged pipe.
PMTs are attached to both sides of the crystal using 3-mm thick optical pad. 
The PTFE flanges are securely fastened to the flanged pipe by pressing the PMT neck, utilizing six stainless steel bolts on each side.
The stainless steel bolts are tightened using a torque wrench to apply same pressure, ensuring and maintaining optical contact between the PMTs and the crystal.
The assembled crystal with PMTs is encapsulated using a copper case, as shown in Fig.~\ref{fig:coppercase}.
The copper case consists of a flanged pipe-shaped encapsulation body and two disk-shaped covers.
When closing the copper covers, six stainless steel bolts on each side are used with a fluorine O-ring.
The three cables (anode, dynode signal readouts, and high voltage input) in each PMT come out through waterproof cable glands (AGM6-3) fastened on the cover using fluorine O-ring and nuts, as shown in Fig.~\ref{fig:cablegland}.
These cable glands are commercial products featuring high waterproofing capabilities, effectively preventing the ingress of LS and external air through the cable exit holes.
The crystal detector assembly is shown in Fig.\,\ref{fig:newdetector}. 
All detector assembly with the new encapsulation design were completed by March 2022. 

We replaced two 4-inch length crystals (NEO-1 and NEO-2~\cite{Choi_2023}) with 8-inch length crystals due to their internal cracks and higher internal backgrounds.
The new 8-inch crystals have similar qualities with other 8-inch length crystals, such as NEO-4 and NEO-5, in terms of light output and internal backgrounds. 
Consequently, the total mass NaI(Tl) crystals  increased from 13.3\,kg to 16.7\,kg.
The crystals were relabeled as DET-1 to DET-6 for operational purposes, and the physics run started in April 2022. During the physics run, PMT-induced noise was not significantly observed in any of the crystals, as shown in Fig.~\ref{fig:meantime2}.

\section{Light Yield Measurement}
\label{ch:LY}

For the measurements of light yield, we utilized the 59.54\,keV gamma peak from the \am source or the internal gamma (46.6 keV) along with X-rays or Auger electron peaks from \pb.  
When the physics run started, we performed an \am source calibration; however, due to strict safety regulations at commercial nuclear power plants, access to the detector was restricted, and periodic \am calibrations were not possible. Therefore, we monitored the light yield using the internal \pb peak located at approximately 49\,keV.
Light yield is obtained in units of PEs/keV by dividing the total charge at the peak of \pb using corresponding energies  with the charge value of single photoelectron (SPE). 
To determine the charge value of SPEs, we collect isolated clusters within the 4.5--7\,$\mu$s range in an 8\,$\mu$s long waveform, minimizing the presence of multiple PE clusters. 

The isolated cluster was identified by finding the local maximum time bin above the discrimination threshold of the pedestal. Starting from the local maximum, the time bin extended both earlier and later until it met the pedestal.   
The isolated cluster time was defined by this local maximum time bin, and its charge was obtained by integrating the charge from the point where it exceeds the pedestal until it returns to the pedestal~\cite{Lee:2005qr}. 
Although search window of 4.5--7\,$\mu$s range reduced the presence of multiple PE clusters, we counted up to four SPE merged clusters for the isolated charge distribution modeling.  

When a scintillation photon strikes the photocathode of the PMT, it can generate a photoelectron, which undergoes amplification through ten stage dynode structure that typically achieve full amplification, comprising normal gain~(NG) processes. 
However, occasionally, a photon may traverse the photocathode and directly strike the first dynode, or a dynode stage may be bypassed due to the back-scattering of the photoelectron at the initial stage of dynode, leading to the production of under-amplified PEs~\cite{SALDANHA201735}.  
This phenomenon is referred to as low gain (LG) process.

\begin{figure}[tbh!]
\begin{center}
\includegraphics[width=0.5\textwidth]{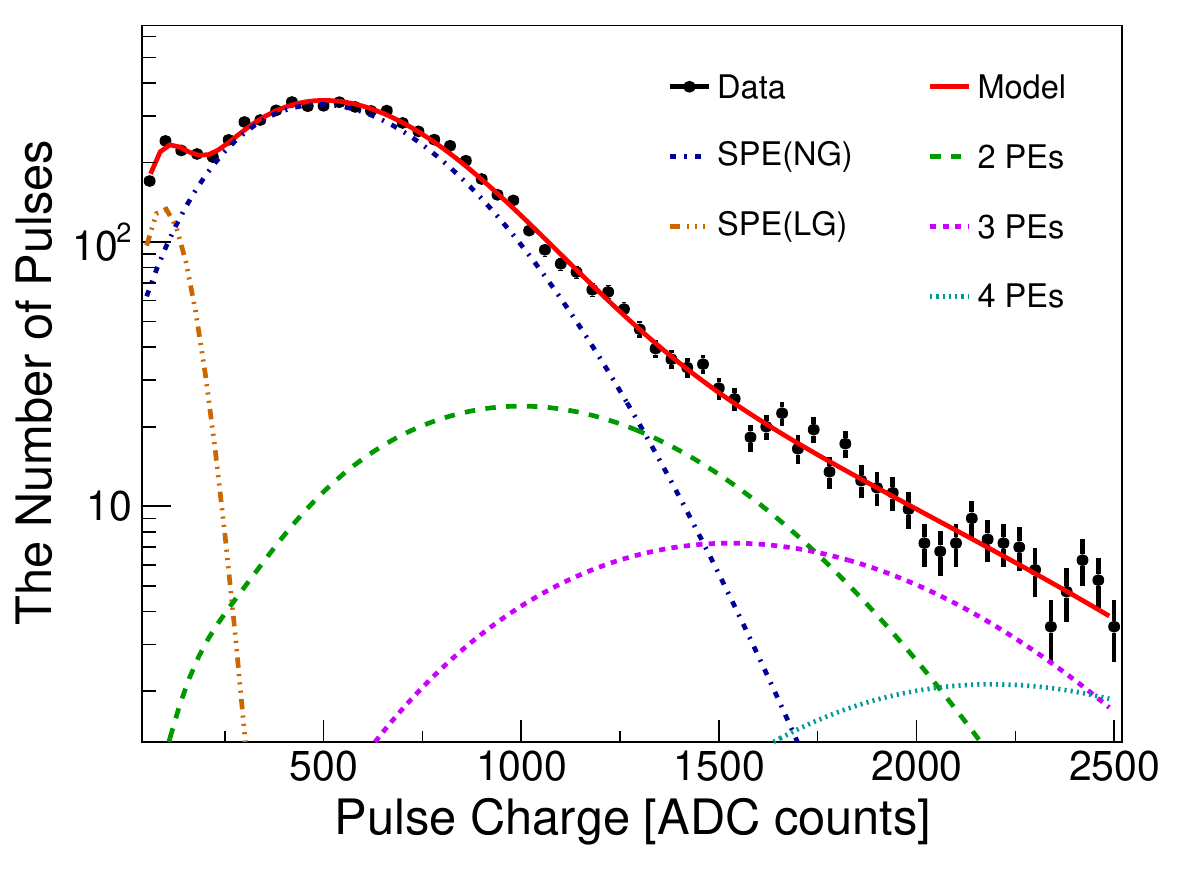} 
\caption{The charge distribution of the isolated clusters in one PMT of DET-4 is fitted by multiple PE and LG SPE. The SPE mean value resulting from the fitting is 572.8\,ADC and the fraction of LG is about 6.8\%.}
\label{fig:LY_4}
\end{center}
\end{figure}

We developed a fit function to account for both LG characteristics and the presence of multiple PE clusters~\cite{Choi:2024ziz}. We can obtain the mean charge of NG SPE and a fraction of LG SPE, as shown in Fig.~\ref{fig:LY_4}. Typical LG fractions vary from 5\,\% to 10\,\%, similar with Ref.~\cite{SALDANHA201735}.

Due to the relatively low charge of LG SPE events, reconstructing all LG PEs from pedestal background is not achievable. With simple height discrimination, approximately an order of 10\,\% of the LG PEs can be isolated depending on the PMTs. If we can improve the LG SPE reconstruction method from the pedestal, which can be enhanced by accounting for the shape of the isolated cluster, we can use information from LG SPE, effectively enhancing the light yield for low-energy signal regions of \cenns.  
For this reason, we account for light yields of the NaI(Tl) crystals for both without LG PEs and with LG PEs, as presented in Table~\ref{table:LightYield_rev}. 

\begin{figure}[!htb]
  \begin{center}
    \includegraphics[width=0.5\textwidth]{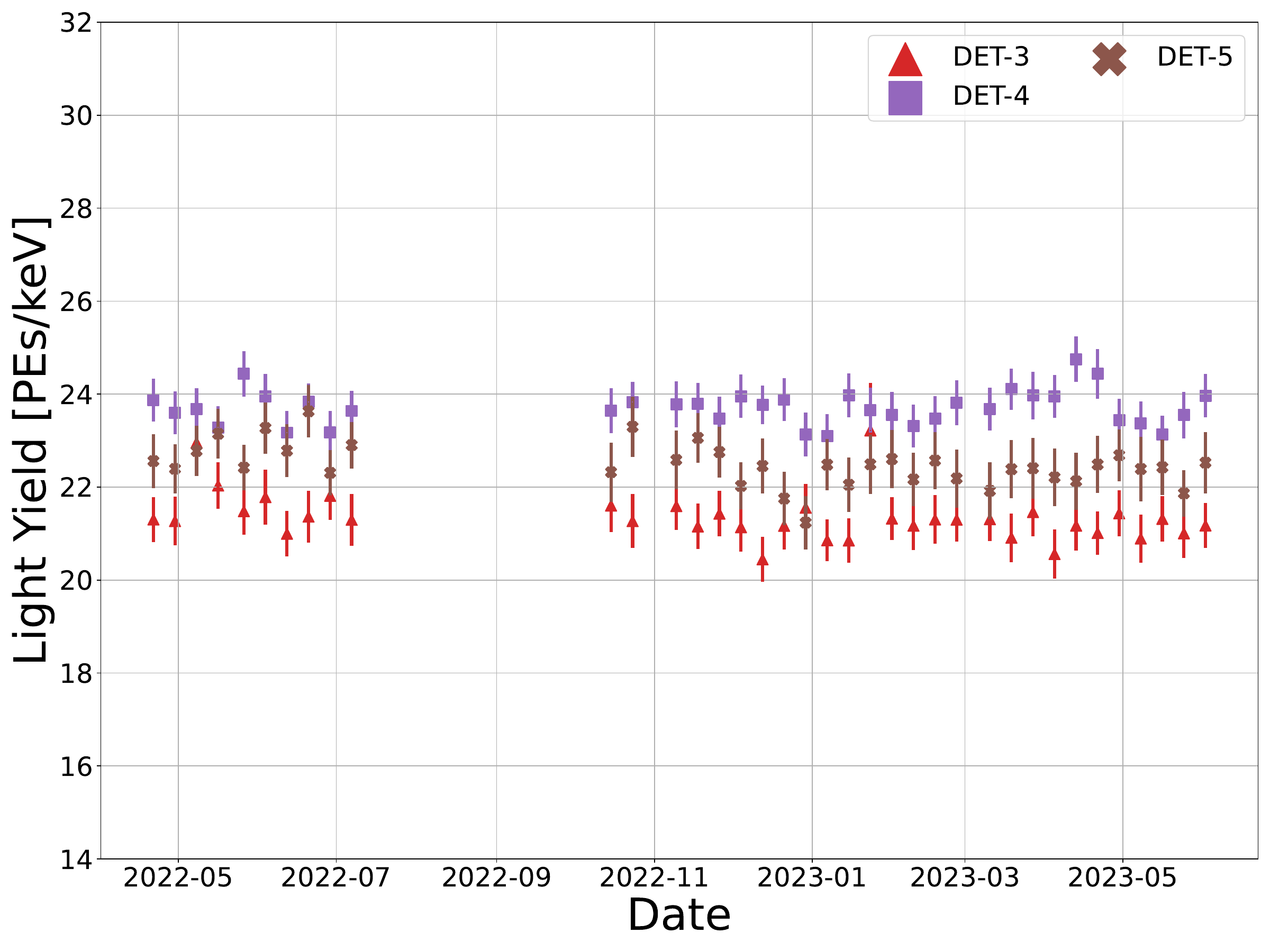}
  \end{center}
  \caption{The light yield monitoring while receiving physics data.
   The colored periods in the figure indicate periods when data collection was interrupted due to power outages and breakdown of high-voltage modules. 
   The light yield of the detectors did not decrease over time during the period of receiving physics data, indicating that the new encapsulation design was stable enough to maintain the light yield. }
  \label{fit:LY_stability}
\end{figure}

During the first 100 hours of the NEON experiment, the NaI(Tl) crystals exhibited light yields mostly exceeding 22\,PEs/keV, even without considering LG SPEs. If we can reconstruct 100\,\% of LG PEs, this yield effectively increase to above 23.5\,PEs/keV. 
The maximum light yield is measured with DET-2 as a 25.6$\pm$1.1\,PEs/keV without LG PEs. 
The light yields of each crystals are monitored using the internal peaks of \pb decay, and stable light yields over one year operation are observed, as one can see in Fig.\,\ref{fit:LY_stability}.
This light yield is well above our initial assumption to achieve more than 3-$\sigma$ sensitivity for \cenns observation~\cite{Choi_2023}.

\begin{table}[bt]
  \begin{center}
  \scalebox{0.9}{
    \begin{tabular}{c c c c c} 
      \hline
      Detector & Mass  & size & Light yield     & Light yield   \\
       &   & &   w/o LG   & w/ LG   \\
         & (kg) & (inch, D$\times$L)    &(PEs/keV)& (PEs/keV) \\\hline
       DET-1 &1.67 &$3\times 4$ & 22.0$\pm$0.4 & 25.3$\pm$0.6\\
       DET-2 &3.34 &$3\times 8$ &25.6$\pm$1.1 & 27.8$\pm$1.4\\
       DET-3 &1.65 &$3\times 4$ &21.8$\pm$0.5 & 23.3$\pm$0.9\\
       DET-4 &3.34 &$3\times 8$ &23.7$\pm$0.4 & 25.4$\pm$0.7\\
       DET-5 &3.35 &$3\times 8$ & 22.4$\pm$0.5 & 23.6$\pm$0.8\\
       DET-6 &3.35 &$3\times 8$ & 25.0$\pm$0.5 & 27.9$\pm$0.7\\
      \hline
    \end{tabular}}
  \end{center}
  \caption{The light yields of the NEON detector were calculated using two methods in the first 100\,hour physics run data. The first method involved using the SPE mean, taking into account only NG SPEs. The second method, on the other hand, consider LG SPEs to account light yield. }
  \label{table:LightYield_rev}
\end{table}

\section{Conclusion}
In this article, we introduced a new NaI(Tl) crystal encapsulation for the NEON experiment.
High light yield up to 25.6$\pm$1.1\,PEs/keV for a large-size NaI(Tl) crystal with a dimension of 3-inch diameter and 8-inch long cylindrical shape was maintained for approximately one year of operation. 
This may allow us to anticipate the observation of \cenns with reactor anti-electron neutrinos from the NEON experiment. 
A similar technique is now being applied for the upgrade of the COSINE-100 experiment to enhance light collection efficiency. This encapsulation technique can be applicable for the next-generation dark matter search and \cenns experiments using the NaI(Tl) crystals.

\section*{Acknowledgments}
We thank the Korea Hydro and Nuclear Power (KHNP) company and particularly acknowledge the help and support provided by the staff members of the Safety and Engineering Support Team of Hanbit Nuclear Power Plant 3.
We thank the IBS Research Solution Center (RSC) for providing high performance computing resources.
This work is supported by the Institute for Basic Science (IBS) under the project code IBS-R016-A1 and NRF-2021R1A2C1013761. 

\bibliography{dm}

\begin{thebibliography}{10}
\expandafter\ifx\csname url\endcsname\relax
  \def\url#1{\texttt{#1}}\fi
\expandafter\ifx\csname urlprefix\endcsname\relax\def\urlprefix{URL }\fi
\expandafter\ifx\csname href\endcsname\relax
  \def\href#1#2{#2} \def\path#1{#1}\fi

\bibitem{PhysRevD.9.1389}
D.~Z. Freedman,
  \href{https://link.aps.org/doi/10.1103/PhysRevD.9.1389}{Coherent effects of a
  weak neutral current}, Phys. Rev. D 9 (1974) 1389.
\newblock \href {http://dx.doi.org/10.1103/PhysRevD.9.1389}
  {\path{doi:10.1103/PhysRevD.9.1389}}.
\newline\urlprefix\url{https://link.aps.org/doi/10.1103/PhysRevD.9.1389}

\bibitem{osti_4289450}
V.~B. Kopeliovich, L.~L. Frankfurt,
  \href{https://www.osti.gov/biblio/4289450}{Isotopic and chiral structure of
  neutral current}, JETP Lett. (USSR) (Engl. Transl.), v. 19, no. 4, pp.
  145-147.
\newline\urlprefix\url{https://www.osti.gov/biblio/4289450}

\bibitem{Akimov:2017ade}
D.~Akimov, et~al., {Observation of Coherent Elastic Neutrino-Nucleus
  Scattering}, Science 357~(6356) (2017) 1123.
\newblock \href {http://arxiv.org/abs/1708.01294} {\path{arXiv:1708.01294}},
  \href {http://dx.doi.org/10.1126/science.aao0990}
  {\path{doi:10.1126/science.aao0990}}.

\bibitem{COHERENT:2020iec}
D.~Akimov, et~al., {First Measurement of Coherent Elastic Neutrino-Nucleus
  Scattering on Argon}, Phys. Rev. Lett. 126~(1) (2021) 012002.
\newblock \href {http://arxiv.org/abs/2003.10630} {\path{arXiv:2003.10630}},
  \href {http://dx.doi.org/10.1103/PhysRevLett.126.012002}
  {\path{doi:10.1103/PhysRevLett.126.012002}}.

\bibitem{COHERENT:Ge}
R.~Bouabid,
  \href{https://indico.cern.ch/event/1342813/contributions/5913875/attachments/2875782/5036212/m7s_rbouabid_upload.pdf}{{First
  Measurement of CE$\nu$NS on Germanium by COHERENT}}, magnificent CE$\nu$NS
  2024 (2024).
\newline\urlprefix\url{https://indico.cern.ch/event/1342813/contributions/5913875/attachments/2875782/5036212/m7s_rbouabid_upload.pdf}

\bibitem{Kerman:2016jqp}
S.~Kerman, et~al., {Coherency in Neutrino-Nucleus Elastic Scattering}, Phys.
  Rev. D 93~(11) (2016) 113006.
\newblock \href {http://arxiv.org/abs/1603.08786} {\path{arXiv:1603.08786}},
  \href {http://dx.doi.org/10.1103/PhysRevD.93.113006}
  {\path{doi:10.1103/PhysRevD.93.113006}}.

\bibitem{Drukier19842295}
A.~Drukier, L.~Stodolsky,
  \href{https://www.scopus.com/inward/record.uri?eid=2-s2.0-33750966650&doi=10.1103%2fPhysRevD.30.2295&partnerID=40&md5=db6a7d821d51b8c47b17be226b568d66}{Principles
  and applications of a neutral-current detector for neutrino physics and
  astronomy}, Phys. Rev. D 30~(11) (1984) 2295.
\newblock \href {http://dx.doi.org/10.1103/PhysRevD.30.2295}
  {\path{doi:10.1103/PhysRevD.30.2295}}.
\newline\urlprefix\url{https://www.scopus.com/inward/record.uri?eid=2-s2.0-33750966650&doi=10.1103%2fPhysRevD.30.2295&partnerID=40&md5=db6a7d821d51b8c47b17be226b568d66}

\bibitem{Krauss:1991ba}
L.~M. Krauss, {Low-energy neutrino detection and precision tests of the
  standard model}, Phys. Lett. B 269 (1991) 407.
\newblock \href {http://dx.doi.org/10.1016/0370-2693(91)90192-S}
  {\path{doi:10.1016/0370-2693(91)90192-S}}.

\bibitem{Patton:2012jr}
K.~Patton, et~al., {Neutrino-nucleus coherent scattering as a probe of neutron
  density distributions}, Phys. Rev. C 86 (2012) 024612.
\newblock \href {http://arxiv.org/abs/1207.0693} {\path{arXiv:1207.0693}},
  \href {http://dx.doi.org/10.1103/PhysRevC.86.024612}
  {\path{doi:10.1103/PhysRevC.86.024612}}.

\bibitem{Formaggio:2011jt}
J.~A. Formaggio, E.~Figueroa-Feliciano, A.~J. Anderson, {Sterile Neutrinos,
  Coherent Scattering and Oscillometry Measurements with Low-temperature
  Bolometers}, Phys. Rev. D 85 (2012) 013009.
\newblock \href {http://arxiv.org/abs/1107.3512} {\path{arXiv:1107.3512}},
  \href {http://dx.doi.org/10.1103/PhysRevD.85.013009}
  {\path{doi:10.1103/PhysRevD.85.013009}}.

\bibitem{deNiverville:2015mwa}
P.~deNiverville, M.~Pospelov, A.~Ritz, {Light new physics in coherent
  neutrino-nucleus scattering experiments}, Phys. Rev. D 92~(9) (2015) 095005.
\newblock \href {http://arxiv.org/abs/1505.07805} {\path{arXiv:1505.07805}},
  \href {http://dx.doi.org/10.1103/PhysRevD.92.095005}
  {\path{doi:10.1103/PhysRevD.92.095005}}.

\bibitem{Dutta:2015nlo}
B.~Dutta, Y.~Gao, R.~Mahapatra, N.~Mirabolfathi, L.~E. Strigari, J.~W. Walker,
  {Sensitivity to oscillation with a sterile fourth generation neutrino from
  ultra-low threshold neutrino-nucleus coherent scattering}, Phys. Rev. D
  94~(9) (2016) 093002.
\newblock \href {http://arxiv.org/abs/1511.02834} {\path{arXiv:1511.02834}},
  \href {http://dx.doi.org/10.1103/PhysRevD.94.093002}
  {\path{doi:10.1103/PhysRevD.94.093002}}.

\bibitem{Dent:2016wcr}
J.~B. Dent, B.~Dutta, S.~Liao, J.~L. Newstead, L.~E. Strigari, J.~W. Walker,
  {Probing light mediators at ultralow threshold energies with coherent elastic
  neutrino-nucleus scattering}, Phys. Rev. D 96~(9) (2017) 095007.
\newblock \href {http://arxiv.org/abs/1612.06350} {\path{arXiv:1612.06350}},
  \href {http://dx.doi.org/10.1103/PhysRevD.96.095007}
  {\path{doi:10.1103/PhysRevD.96.095007}}.

\bibitem{Kosmas:2017zbh}
T.~S. Kosmas, et~al., {Probing light sterile neutrino signatures at reactor and
  Spallation Neutron Source neutrino experiments}, Phys. Rev. D 96~(6) (2017)
  063013.
\newblock \href {http://arxiv.org/abs/1703.00054} {\path{arXiv:1703.00054}},
  \href {http://dx.doi.org/10.1103/PhysRevD.96.063013}
  {\path{doi:10.1103/PhysRevD.96.063013}}.

\bibitem{Liao:2017uzy}
J.~Liao, D.~Marfatia, {COHERENT constraints on nonstandard neutrino
  interactions}, Phys. Lett. B 775 (2017) 54.
\newblock \href {http://arxiv.org/abs/1708.04255} {\path{arXiv:1708.04255}},
  \href {http://dx.doi.org/10.1016/j.physletb.2017.10.046}
  {\path{doi:10.1016/j.physletb.2017.10.046}}.

\bibitem{Farzan:2018gtr}
Y.~Farzan, M.~Lindner, W.~Rodejohann, X.-J. Xu, {Probing neutrino coupling to a
  light scalar with coherent neutrino scattering}, JHEP 05 (2018) 066.
\newblock \href {http://arxiv.org/abs/1802.05171} {\path{arXiv:1802.05171}},
  \href {http://dx.doi.org/10.1007/JHEP05(2018)066}
  {\path{doi:10.1007/JHEP05(2018)066}}.

\bibitem{Dev:2019anc}
P.~Bhupal~Dev, et~al., {Neutrino Non-Standard Interactions: A Status Report} 2
  (2019) 001.
\newblock \href {http://arxiv.org/abs/1907.00991} {\path{arXiv:1907.00991}},
  \href {http://dx.doi.org/10.21468/SciPostPhysProc.2.001}
  {\path{doi:10.21468/SciPostPhysProc.2.001}}.

\bibitem{Aguilar-Arevalo:2019jlr}
A.~Aguilar-Arevalo, et~al., {Exploring low-energy neutrino physics with the
  Coherent Neutrino Nucleus Interaction Experiment}, Phys. Rev. D 100~(9)
  (2019) 092005.
\newblock \href {http://arxiv.org/abs/1906.02200} {\path{arXiv:1906.02200}},
  \href {http://dx.doi.org/10.1103/PhysRevD.100.092005}
  {\path{doi:10.1103/PhysRevD.100.092005}}.

\bibitem{Janka:2017vlw}
H.~T. Janka, {Neutrino Emission from Supernovae}\href
  {http://arxiv.org/abs/1702.08713} {\path{arXiv:1702.08713}}, \href
  {http://dx.doi.org/10.1007/978-3-319-21846-5_4}
  {\path{doi:10.1007/978-3-319-21846-5_4}}.

\bibitem{Cogswell2016}
B.~K. Cogswell, P.~Huber, {Detection of Breeding Blankets Using Antineutrinos},
  Science \& Global Security 24~(2) (2016) 114--130.
\newblock \href {http://dx.doi.org/10.1080/08929882.2016.1184531}
  {\path{doi:10.1080/08929882.2016.1184531}}.

\bibitem{RevModPhys.92.011003}
A.~Bernstein, et~al.,
  \href{https://link.aps.org/doi/10.1103/RevModPhys.92.011003}{Colloquium:
  Neutrino detectors as tools for nuclear security}, Rev. Mod. Phys. 92 (2020)
  011003.
\newblock \href {http://dx.doi.org/10.1103/RevModPhys.92.011003}
  {\path{doi:10.1103/RevModPhys.92.011003}}.
\newline\urlprefix\url{https://link.aps.org/doi/10.1103/RevModPhys.92.011003}

\bibitem{CONUS:2020skt}
H.~Bonet, et~al., {Constraints on Elastic Neutrino Nucleus Scattering in the
  Fully Coherent Regime from the CONUS Experiment}, Phys. Rev. Lett. 126~(4)
  (2021) 041804.
\newblock \href {http://arxiv.org/abs/2011.00210} {\path{arXiv:2011.00210}},
  \href {http://dx.doi.org/10.1103/PhysRevLett.126.041804}
  {\path{doi:10.1103/PhysRevLett.126.041804}}.

\bibitem{Angloher:2019flc}
G.~Angloher, et~al., {Exploring CE$\nu$NS with NUCLEUS at the Chooz nuclear
  power plant}, Eur. Phys. J. C 79~(12) (2019) 1018.
\newblock \href {http://arxiv.org/abs/1905.10258} {\path{arXiv:1905.10258}},
  \href {http://dx.doi.org/10.1140/epjc/s10052-019-7454-4}
  {\path{doi:10.1140/epjc/s10052-019-7454-4}}.

\bibitem{Aguilar-Arevalo:2016khx}
A.~Aguilar-Arevalo, et~al., {The CONNIE experiment}, J. Phys. Conf. Ser.
  761~(1) (2016) 012057.
\newblock \href {http://arxiv.org/abs/1608.01565} {\path{arXiv:1608.01565}},
  \href {http://dx.doi.org/10.1088/1742-6596/761/1/012057}
  {\path{doi:10.1088/1742-6596/761/1/012057}}.

\bibitem{Belov:2015ufh}
V.~Belov, et~al., {The $\nu$GeN experiment at the Kalinin Nuclear Power Plant},
  JINST 10~(12) (2015) P12011.
\newblock \href {http://dx.doi.org/10.1088/1748-0221/10/12/P12011}
  {\path{doi:10.1088/1748-0221/10/12/P12011}}.

\bibitem{Choi_2023}
J.~J. Choi, et~al.,
  \href{http://dx.doi.org/10.1140/epjc/s10052-023-11352-x}{Exploring coherent
  elastic neutrino-nucleus scattering using reactor electron antineutrinos in
  the neon experiment}, Eur. Phys. J. C 83~(3).
\newblock \href {http://dx.doi.org/10.1140/epjc/s10052-023-11352-x}
  {\path{doi:10.1140/epjc/s10052-023-11352-x}}.
\newline\urlprefix\url{http://dx.doi.org/10.1140/epjc/s10052-023-11352-x}

\bibitem{CHOI2020164556}
J.~J. Choi, et~al., {Improving the light collection using a new NaI(Tl) crystal
  encapsulation}, Nucl. Instrum. Meth. A 981 (2020) 164556.
\newblock \href {http://dx.doi.org/https://doi.org/10.1016/j.nima.2020.164556}
  {\path{doi:https://doi.org/10.1016/j.nima.2020.164556}}.

\bibitem{Adhikari:2017esn}
G.~Adhikari, et~al., {Initial Performance of the COSINE-100 Experiment}, Eur.
  Phys. J. C 78~(2) (2018) 107.
\newblock \href {http://dx.doi.org/10.1140/epjc/s10052-018-5590-x}
  {\path{doi:10.1140/epjc/s10052-018-5590-x}}.

\bibitem{Adhikari:2020asl}
G.~Adhikari, et~al., {The COSINE-100 liquid scintillator veto system}, Nucl.
  Instrum. Meth. A 1006 (2021) 165431.
\newblock \href {http://arxiv.org/abs/2004.03463} {\path{arXiv:2004.03463}},
  \href {http://dx.doi.org/10.1016/j.nima.2021.165431}
  {\path{doi:10.1016/j.nima.2021.165431}}.

\bibitem{kwkim15}
K.~W. Kim, et~al., {Tests on NaI(Tl) crystals for WIMP search at the Yangyang
  Underground Laboratory}, Astropart. Phys. 62 (2015) 249.

\bibitem{Lee:2005qr}
H.~S. Lee, et~al., {First limit on wimp cross section with low background
  csi(tl) crystal detector}, Phys. Lett. B 633 (2006) 201.
\newblock \href {http://dx.doi.org/10.1016/j.physletb.2005.12.035}
  {\path{doi:10.1016/j.physletb.2005.12.035}}.

\bibitem{SALDANHA201735}
R.~Saldanha, et~al.,
  \href{https://www.sciencedirect.com/science/article/pii/S016890021730311X}{Model
  independent approach to the single photoelectron calibration of
  photomultiplier tubes}, Nucl. Instrum. Meth. A 863 (2017) 35--46.
\newblock \href {http://dx.doi.org/https://doi.org/10.1016/j.nima.2017.02.086}
  {\path{doi:https://doi.org/10.1016/j.nima.2017.02.086}}.
\newline\urlprefix\url{https://www.sciencedirect.com/science/article/pii/S016890021730311X}

\bibitem{Choi:2024ziz}
J.~J. Choi, et~al., {Waveform Simulation for Scintillation Characteristics of
  NaI(Tl) Crystal. }\href {http://arxiv.org/abs/2402.17125}
  {\path{arXiv:2402.17125}}.

\end{thebibliography}
\end{document}